\documentclass[12pt]{article}
\usepackage{graphicx}
\DeclareGraphicsExtensions{.eps}
\begin{document}
\pagestyle{plain}
\pagenumbering{arabic}
\flushbottom
\begin{titlepage}
\title{A curvature dependent bound for entanglement change in classically chaotic systems}
\author{Demetris P.K. Ghikas,\thanks{ghikas@physics.upatras.gr, Tel: +302610-997460, FAX: +302610-997617}
\and George Stamatiou
\and Department of Physics
University of Patras,
Patras 26500, Greece}
\maketitle
\begin{abstract}
Using the Calogero-Moser model and the Nakamura equations for a multi-partite quantum system, we prove an inequality between the mean bi-partite entanglement rate of change under the variation of a critical parameter and the level-curvature. This provides an upper bound for the rate of production or distraction of entanglement induced dynamically. We then investigate the dependence of the upper bound on the degree of chaos of the system, which in turn, through the inequality, gives a measure of the stability of the entangled state. Our analytical results are supported with extensive numerical calculations. 
\newline\      
PACS : 03.67Mn, 05.45Ac, 05.45Pq
\newline\   
KEY WORDS : Entanglement, Quantum Chaos, Calogero-Moser Model,
\newline\   
upper bounds. 
\end{abstract}
\end{titlepage}
\newpage
\noindent
\,1.\, INTRODUCTION
\vspace{1 cm}\\
\indent To store , process or teleport quantum information we need real, physical quantum systems which live outside the "protected environment " of the quantum mechanics text books, and beyond the safe bounds of linearity \cite{nielsen,blanchard}. These Quantum Information Devises (Q.I.D.) are always open quantum systems and complex enough that the effects of non-linearity cannot be ignored. Given the fact that the prime and most relevant physical issue for QID is the support of quantum information, the main interest of most analytical, numerical and experimental work  currently done is about the production and protection of entanglement, which is the manifestation of non-classical information in quantum systems. Since the physical devises are open systems and may operate in a non-linear regime, it is a natural question to be asked whether the dynamics of production or distraction of entanglement is related to the dynamical properties of the corresponding classical systems. There are many works dealing with this problem, namely , the connection between entanglement and quantum chaos,\cite{bell,mona,fuji,ghos,bandy1,bandy2,wang,jac,znid,pine}. While it seems that there are some controversial results about the distractive or constructive influence of classical dynamics on entanglement , one can classify and understand the conditions for one or the other behavior. The main problem with all these results is that they are model and initial state dependent. A better approach would use the Random Matrix ideas to obtain general results \cite{gorin}. Thus a general result in terms of inequalities and bounds can provide an insight for the relevant effects. Here we prove such an inequality which bounds the rate of change of mean bi-partite entanglement of a multi-partite system when a critical parameter is changed. The bound is proportional to  the square root of level-curvature of the particular state of the system.Then using the level-curvature distribution function for different ensembles we find some optimality conditions which we believe are relevant for the design of QID. We made extensive numerical tests of the inequality using three different models. The inequality is satisfied  strictly. Now , the established inequality can be applied to investigate the stability properties of the entangled state w.r.t. the degree of chaos in the system. The latter is quantified with the dependence of the upper bound on a critical parameter which controls the variation of the system from the regular to the chaotic regimes. We do that for two physically different systems. The first is described by a Hamiltonian which is the sum of a random matix with Poisson level spacing distribution and a random matrix with GOE distribution, \cite{guhr}. The second system is a one dimensional spin chain with a random distribution of defects \cite{santos}. First we confirm  the transition from regularity to chaos by plotting the dependence of the parameter characterizing the degree of chaos vs the critical parameter of the dynamics. Then we compute the dependence of the bound on both the chaos parameter and the dynamics parameter. It is obvious from these plots , and the inequality , that the variation of the mean bi-partite entanglement is bounded more strictly when the investigated systems are more regular. We believe that this is an indication of the usefulness of the inequality which has a model independent validity.   
\newline
\newpage
\,2.\, CHANGE OF ENTANGLEMENT AND AVOIDED CROSSINGS.
\vspace{1 cm}\\
\indent
We start from the autonomous non-integrable system with Hamiltonian 
\begin{equation}
H\,=\,H_{0}\,+\,\tau V
\end{equation}
where $H_{0}$ is integrable and $\tau V$ a non-integrable pertubation with $\tau $ the perturbation parameter. Let the eigenvalue equation for each $\tau$ be discrete and non-degenerate
\begin{equation}
H(\tau)|n(\tau)>\,=\,\epsilon_{n}(\tau)|n(\tau)>
\end{equation}
 Then considering the parameter $\tau$ as a 'time' the dynamical system $(\epsilon_{n}(\tau),p_{n}(\tau))$ satisfies the equations \cite{nakabook,naka,taka1,taka2}
 \begin{equation}
 \frac{d\epsilon_{n}(\tau)}{d\tau}\,=\,p_{n}
 \end{equation} 
 \begin{equation}
 \frac{dp_{n}(\tau)}{d\tau}\,=\,2\sum_{m(\ne n)}V_{nm}(\tau)V_{mn}(\tau)(\epsilon_{n}(\tau)-\epsilon_{m}(\tau))^{-1}
 \end{equation}
 \begin{equation}
 \frac{d|n(\tau)>}{d\tau}\,=\,\sum_{m(\ne    n)}|m(\tau)>V_{mn}(\tau)(\epsilon_{n}(\tau)-\epsilon_{m}(\tau))^{-1}
 \end{equation}
 \begin{equation}
 \frac{d<n(\tau)|}{d\tau}\,=\,\sum_{m(\ne    n)}<m(\tau)|V_{mn}(\tau)(\epsilon_{n}(\tau)-\epsilon_{m}(\tau))^{-1}
 \end{equation}
 where $p_{n}(\tau)\,\equiv\,V_{nn}\,=\,<n(\tau)|V|n(\tau)>$.
 These equations can be transformed into an integrable Hamiltonian system, the Calogero-Moser N-particle system in 1+1 dimensions.
 Now we suppose that we have a bi-partite system with subsystems A and B and the total system dynamics given by equation (1). Let the pure state
 \begin{equation}
 \rho^{n}_{AB}(\tau)\,=\,|n(\tau)><n(\tau)|
 \end{equation}
 Using the linear entropy $E_{L}$ as a measure of entanglement we analyze its dependence on the parameter $\tau$.
 We have
 \begin{equation}
 E_{L}\,=\,1\,-\,tr((\rho^{n}_{A})^{2})
 \end{equation}
 where
 \begin{equation}
 \rho^{n}_{A}\,=\,tr_{B}(\rho^{n}_{AB})
 \end{equation}
 We want to compute the derivative  $dE_{L}/d\tau$.
 We have
 \begin{equation} \frac{\partial}{\partial\tau}tr[(\rho^{n}_{A}(\tau))^{2}]\,=\,2tr[tr_{B}(\rho^{n}_{AB})tr_{B}[\frac{\partial}{\partial\tau}(\rho^{n}_{AB})]]
 \end{equation}
 Now
 \begin{eqnarray*} <k(\tau)|\frac{\partial}{\partial\tau}\rho^{n}_{AB}(\tau)|l(\tau)>\\=<k(\tau)|(\frac{\partial}{\partial\tau}|n(\tau)>)<n(\tau)|l(\tau)>&+&<k(\tau)|n(\tau)>(\frac{\partial}{\partial\tau}<n(\tau)|)|l(\tau)>\\&=&<k(\tau)|(\frac{\partial}{\partial\tau}|n(\tau)>)\delta_{nl}+\delta_{kn}(\frac{\partial}{\partial\tau}<n(\tau)|)|l(\tau)>
 \end{eqnarray*}
 and this , using Nakamura's equations above, is equal to
\begin{equation} \gamma^{n}_{kl}(\tau)\,\equiv\,\delta_{nl}(1-\delta_{nk})\frac{V_{kn}}{\epsilon_{n}-\epsilon_{k}}+\delta_{nl}(1-\delta_{kn})\frac{V_{nl}}{\epsilon_{n}-\epsilon_{l}}
\end{equation}
Thus
\begin{equation}
\frac{\partial}{\partial\tau}\rho^{n}_{AB}(\tau )\,=\,\sum_{kl}\gamma^{n}_{kl}|k(\tau)><l(\tau)|
\end{equation}
and finally
\begin{equation}
\frac{\partial}{\partial\tau}E^{n}_{L}\,=\,-2\sum_{kl}\gamma^{n}_{kl}tr[tr_{B}(|n(\tau)><n(\tau)|)tr_{B}(|k(\tau)><l(\tau)|)]
\end{equation}
or
\begin{equation}
\frac{\partial}{\partial\tau}E^{n}_{L}\,=\,-2\sum_{kl}\gamma^{n}_{kl}A^{n}_{kl}(\tau)
\end{equation}
with
\begin{equation}
A^{n}_{kl}(\tau)\,=\,tr[tr_{B}(|n(\tau)><n(\tau)|)tr_{B}(|k(\tau)><l(\tau)|)]
\end{equation}
We observe here that, while $A^{n}_{kl}(\tau) $ is a bounded quantity, it can be proved that $\gamma^{n}_{kl}$ has first order poles at the avoided crossings, a fact we avoid assuming non-degeneracy of the spectrum. But if the system is classically chaotic we have many avoided crossings, and this leads to the result that for the corresponding states, at these critical values of the parameter $\tau$ we expect fast change of the entaglement under a small change of the parameter. 
 
\noindent
\newline
\newpage
\,3.\, CHANGE OF ENTANGLEMENT AND LEVEL-CURVATURE.
\vspace{1 cm}\\
\indent
Here we are going to show that the rate of entanglement change at the particular value of the parameter $\tau$ is proportional to the square root of the level-curvature, under the assumption that for this value an avoided crossing of two levels gives so a small energy gap that only these two levels contribute to the sums above. We call these levels 0 and 1. First we write equation (14) as
\begin{equation}
\frac{\partial E^{n}_{L}}{\partial\tau}\,=\,-2\sum_{k\ne n}\frac{V_{kn}A^{n}_{kn}+V_{nk}A^{n}_{nk}}{\epsilon_{n}-\epsilon_{k}}.
\end{equation}
From Nakamura's equations we have for the level-curvature 
\begin{equation}
K_{n}\,\equiv\,\frac{\partial^{2}}{\partial\tau^{2}}E^{n}_{L}\,=\,\sum_{m\ne n}\frac{|V_{nm}|^{2}}{\epsilon_{n}-\epsilon_{m}}
\end{equation}
Under the assumption stated above we have
\begin{equation}
\frac{\partial E^{0}_{L}}{\partial\tau}\,\approx \,-2\frac{V_{10}A^{0}_{10}+V_{01}A^{0}_{01}}{\epsilon_{0}-\epsilon_{1}}
\end{equation}
and similarly
\begin{equation}
\frac{\partial E^{1}_{L}}{\partial\tau}\,\approx \,-2\frac{V_{01}A^{1}_{01}+V_{10}A^{1}_{10}}{\epsilon_{1}-\epsilon_{0}}
\end{equation}
For the curvatures we have
\begin{equation}
K_{0}\,=\,2\frac{|V_{01}|^{2}}{\epsilon_{0}-\epsilon_{1}}\,< \,0
\end{equation}
and
\begin{equation}
K_{1}\,=\,2\frac{|V_{01}|^{2}}{\epsilon_{1}-\epsilon_{0}}\,> \,0
\end{equation}
that is under the assumption we have $K_{0}\,=\,-K_{1}$. We then obtain
\begin{equation}
\frac{\partial E^{0}_{L}}{\partial\tau}\,\approx \,-2\frac{V_{10}A^{0}_{10}+V_{01}A^{0}_{01}}{|V_{01}|^{2}}K_{0}
\end{equation}
and eliminating the potential we get
\begin{equation}
\frac{\partial E^{0}_{L}}{\partial\tau}\,\approx \,\frac{2\sqrt{2} Re[A^{0}_{10}e^{i\phi}]}{\sqrt{\epsilon_{1}-\epsilon_{0}}}\sqrt{|K_{0}|}
\end{equation}
and similarly for the level 1. $\phi$ is a phase related to the potential matrix elements. We may note here that from the last two equations we have near the avoided crossings 
\begin{equation}
\frac{\Delta E^{1}_{L}}{\Delta E^{0}_{L}}\,=\,\frac{Re [ A^{1}_{01}e^{i\phi}]}{ Re [A^{0}_{10}e^{i\phi}]}
\end{equation}
a relation between the entanglement changes of the two levels.  
\noindent
\newline
\vspace{1 cm}\\
\,4.\, GENERALIZATION TO AN N-QUBIT SYSTEM.
\vspace{1 cm}\\
\indent
For an N-qubit system we use as a measure the mean bi-partite entanglement for the level n given by
\begin{equation}
Q^{n}\,=\,2-\frac{2}{N}\sum_{j=1}^{N}tr[\rho^{2}_{j}]
\end{equation}
where $\rho_{j}\,=\,tr_{ALL\ne j}(|n><n|)$
Motivated by the above results one may conjecture that there is a similar relation between the rate of change of the average entanglement and some functional of level-curvature. It turns out that it is straightforward to obtain an inequality for the ground state in the form
\begin{equation}
 |\frac{\partial Q^{0}}{\partial\tau}|\,\leq\,b\sqrt{|K_{0}|}
\end{equation} 
Following the same procedure as that for the bi-partite entanglement rate we get the expression
\begin{equation}
\frac{\partial Q^{0}}{\partial\tau}\,=\,\frac{8}{N}\sum_{k=1}^{2^{N}-1}\frac{Re[V_{0k}A^{0}_{0k}]}{\epsilon_{k}-\epsilon_{0}}
\end{equation}
where here we have $-N \leq A^{0}_{kl}\leq +N$
We prove the inequality by taking the maximum of the r.h.s of (26). 
We have
\begin{eqnarray}
max\frac{\partial Q^{0}}{\partial\tau}=8\sum_{k=1}^{2^{N}-1}\frac {Re|V_{0k}|}{\epsilon_{k}-\epsilon_{0}} \leq 8\sqrt{\sum_{k=1}^{2^{N}-1}\frac {Re^{2}V_{0k}}{\epsilon_{k}-\epsilon_{0}}}\sqrt{\sum_{k=1}^{2^{N}-1}\frac {1}{\epsilon_{k}-\epsilon_{0}}}\\ \leq 8\sqrt{\sum_{k=1}^{2^{N}-1}\frac {|V_{0k}|^{2}}{\epsilon_{k}-\epsilon_{0}}}\sqrt{\sum_{k=1}^{2^{N}-1}\frac {1}{\epsilon_{k}-\epsilon_{0}}}\,=\,b\sqrt{|K_{0}|}
\end{eqnarray}
where
\begin{equation}
b\,=\,8\sqrt{\sum_{k=1}^{2^{N}-1}\frac {1}{\epsilon_{k}-\epsilon_{0}}}
\end{equation}
Our numerical investigation, using Random Matrix Theory, indicated that   
this value of b is not optimal for the inequality. We may obtain a smaller bound under certain statistical assumptions, but still without involving particular properties of the dynamics. The bound is , at least for our numerical results , amazingly good. The derivation is as follows.
In (26) $A^{0}_{kl}$ is a sum of N terms for the N-qubit system
\begin{equation}
A^{0}_{kl}\,=\,\sum_{j=1}^{N}A^{0}_{j;kl}
\end{equation}
Now we assume that these N terms are independent random varables uniformly distributed in the interval [-a,a] with zero mean. From this it follows for the standard deviation $\sigma(A^{0}_{kl})\,=\,\frac{a\sqrt{N}}{\sqrt{3}}$. Our assumption is to use for the derivation of the inequality the plausible bounds
\begin{equation}
-\sigma(A^{0}_{kl})\leq A^{0}_{kl} \leq  \sigma(A^{0}_{kl})
\end{equation}
This lead to the expression
\begin{equation}
b'\,=\,\frac{8a}{\sqrt{3N}}\sqrt{\sum_{k=1}^{2^{N}-1}\frac {1}{\epsilon_{k}-\epsilon_{0}}}
\end{equation}
Now it is natural to suppose that $a\,=\,1/N $, but we get very good results taking the value $a\,=\,1/2^{N}$. We do not have an obvious answer for that.
\noindent
\newline
\vspace{1 cm}\\
\,5.\, ENTANGLEMENT VARIATION AND THE DEGREE OF CHAOS OF THE SYSTEM.
\vspace{1 cm}\\
\indent
\,5.1\, Dependence of the bound on the degree of chaos.
\vspace{1 cm}\\
\indent
Inequality (26) provides a general bound for the rate of change of mean bi-partite entanglement. One may view this inequality as a criterion for the stability of entanglement under small changes of a critical parameter . The smaller the upper bound the more restricted the change of the entanglement. On the other hand a bigger upper bound cannot exclude small variation of the entanglement . For this we would need a lower bound. But still, it is interesting to investigate the dependence of the bound on the degree of chaos. Namely, it would be of some value to see whether integrability or chaos restricts more or less the possible change of entanglement or better whether there is an increased stability of entanglement which depends on the dynamical properties of the system. We use as a characterization of the dynamics the level statistic parameter \cite{guhr}
\begin{equation}
\gamma\,=\,\frac{\int\limits_{0}^{s_{0}}[P(s)-P_{WD}]ds}{\int\limits_{0}^{s_{0}}[P_{P}(s)-P_{WD}]ds}
\end{equation}    
where $s_{0}\approx 0.472$, which is  the common point of Poisson $P_{P}(s)\,=\,exp(-s)$ and Wigner-Dyson $P_{WD}(s)\,=\,\frac{\pi s}{2}exp(-\pi s^{2}/4)$ distributions, and P(s) the level density sistribution of the system. This parameter provides a measure of the degree of chaos, or better of relative distance of the level density from either the Poisson or the Wigner-Dyson distributions.  We use two physically different models to analyze this question. The first is given by a Hamiltonian which is the sum of two random matrices, one with Poisson level statistics and the other with GOE, \cite{guhr}. This superposition is a rotation which depends on an angle $\theta$ which can be varied to give a continuous set of Hamiltonians, from a pure integrable to a pure chaotic one, in the RMT framework. We first study the dependence of $\gamma$ on $\theta$ to establish the variation of the behavior of the system from the regular to the chaotic regimes. Then we make the plots of the dependence of the upper bound parameter b on both $\gamma$ and $\theta$. These plots give a very prominent dependence of the bound on the dynamics of the system. We comment on this in more detail bellow. The second system is a one-dimensional chain of pair-wise coupled spins perturbed by a random distribution of defect sites \cite{santos}. These defects are introduced as differences in the two-level energies on these sites. Here the critical parameter for the transistion from the regular to the chaotic regime is the two-point correlation parameter d of the normally distributed random defects. We plot again the dependence of $\gamma$ on d to establish the behavior of the system. Then we plot the dependence of b, the parameter of the bound, on both $\gamma$ and d. We observe the same behavior of the bound as with the first system. Again we comment in more detail bellow. 
\vspace{1 cm}\\
\indent    
\,5.2\, Entanglement variation for different ensembles.
\vspace{1 cm}\\
\indent
We may apply the derived inequality to investigate the sensitivity of entanglement under small changes of the parameter $\tau$ due to appriximations or environmental changes. The maximum change of the entanglement would be
\begin{equation}
\Delta Q\,=\, |\frac{\partial Q^{0}}{\partial\tau}|\Delta \tau\,=\,b\sqrt{K_{0}}\Delta\tau
\end{equation}
Using this we may observe the differences in the influence between various symmetry ensembles. To do this we calculate the average of the entanglement change using the distribution functions of the level-curvatures, that is we compute \cite{gas,saher,taka3,zakr}
\begin{equation}
<\Delta Q>\,=\,b<\sqrt{K_{0}}>\Delta \tau
\end{equation}  
We have
\begin{equation}
<\sqrt{|K|}>_{GOE}\,\approx \,0.84\sqrt{\gamma_{1}}
\end{equation}
\begin{equation}
<\sqrt{|K|}>_{GUE}\,\approx \,\sqrt{2\gamma_{2}}
\end{equation}

\begin{equation}
<\sqrt{|K|}>_{GSE}\,\approx \,0.6\sqrt{\gamma_{4}}
\end{equation}  
where
\begin{equation}
\gamma_{\nu}\,=\,\nu A
\end{equation}
where A is an appropriate constant related to the ensemble density \cite{zakr}. Then assuming a common value for b, this gives an estimate of the relative merit of the ensembles
\begin{equation}
\frac{\Delta Q_{GOE}}{\Delta Q_{GUE}}\,\approx \, 0.84\,\,,\,\,\frac{\Delta Q_{GOE}}{\Delta Q_{GSE}}\,\approx \, 0.7
\end{equation}
Of course , both factors in the r.h.s. of the inequality depend on the dynamics, as it was stressed above, but still this result could be used as an indication of the relative merit for different ensembles. In the designing strategy for devices which employ the property of entanglement the most serious problem is the fragiliry of resource. We believe that  qualitative properties as the one in (41) could be a guide in this area. We currently investigate this point.  
\noindent
\newline
\vspace{1 cm}\\
\,6.\, MODELS AND NUMERICAL RESULTS.
\vspace{1 cm}\\
\indent
\,6.1\, Testing the inequality.
\vspace{1 cm}\\
\indent
We have checked the inequality (26) numerically choosing three different categories of  models. All are of the form of (1) .
\newline
\newline
\indent
Model A
\newline
\indent We take here 3  qubits with pairwise interactions and perturbed globally by a potential V. For $H_{0}$ we choose
\newline
\begin{equation}
H_{0}\,=\,\sum_{j=1}^{3}a_{j}\sigma_{zj}\,+\,\lambda \sum_{i<j}\vec{\sigma_{i}}\cdot \vec{\sigma_{j}}
\end{equation}  
where $a_{j}$ and $\lambda$ are arbitrary constants. Our choice gives b=9.849476 and  b'=0.410395. V is Hermitian with random matrix elements. 
\newline
\newline
\indent
Model B
\newline
\indent $H_{0}$ is an arbitrary Hermitian matrix and V as in Model A. For 2 qubits we have b=11.66033 , b'=1.190078 and for 3 qubits b=20.57092 , b'=0.8571215 and for 6 qubits b=38.85935 , b'=0.1431131.
\newline
\newline
\indent
Model C
\newline
\indent 2 qubits with $H_{0}$ as in Model B. Two choices for V, from GUE and GOE.
\newline
\newline
\indent
For all models we computed the two sides of the inequality (26), for the given choice of $H_{0}$ and for 3000 choices of V. For each pair of values we put a point in the graph of $|\frac{\partial Q^{0}}{\partial\tau}|$ versus $\sqrt{|K_{0}|}$. In the same graph we draw the line $y\,=\,b'\sqrt{|K_{0}|} $. For all models we observe that all points, but for few, are bellow this line. These outliers are due to the fact that we have estimated b' statistically. To make the results more obvious, for each case we draw a second graph for the difference of the l.h.s minus r.h.s of inequality (25). i.e.   $\delta\,=\,|\frac{\partial Q^{0}}{\partial\tau}|_{numerical} - b\sqrt{|K_{0}|}$ . We call $\delta$ 'saturation index' since it indicates how far we are from the saturation of the inquality.
\noindent
\newline
\begin{figure}[htp]
	\centering
\scalebox{0.60}{\includegraphics{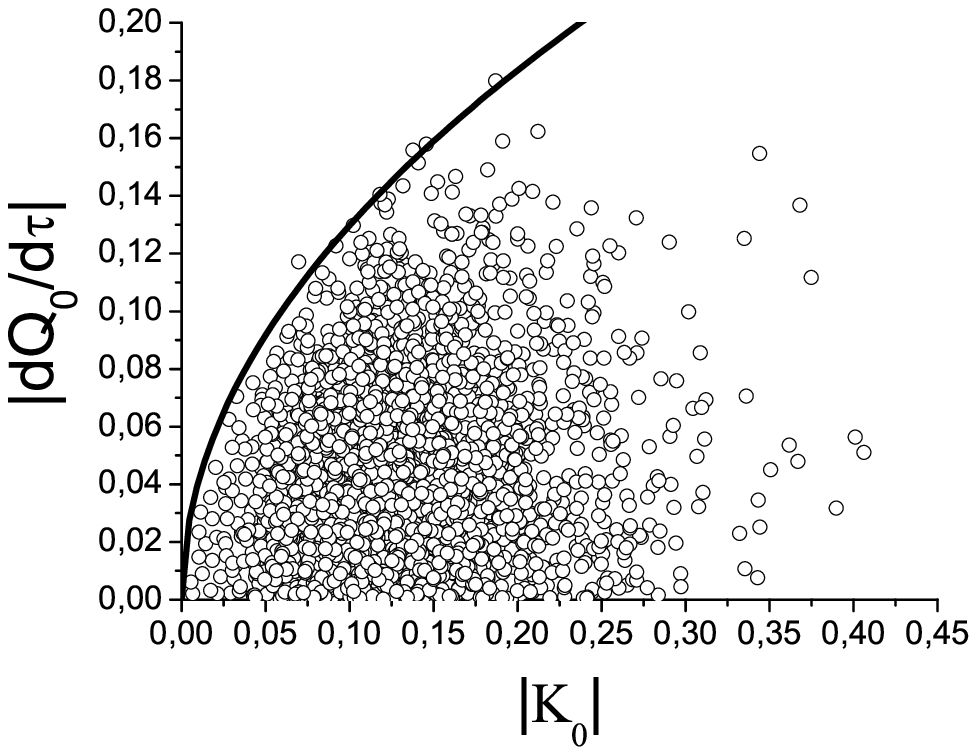}}
\scalebox{0.60}{\includegraphics{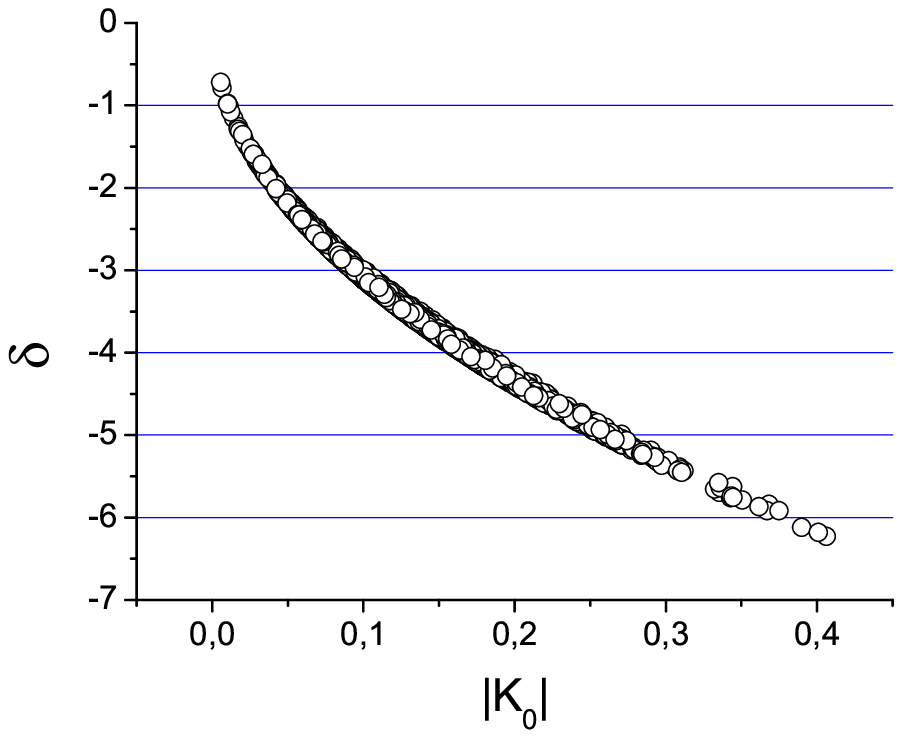}}
	\caption{Model A: 3 qubits. Numerical test of (26). Left : distribution of 3000 random points. Right: dependence of the saturation index on curvature  }
	\label{fig:model A}
\end{figure}
 
 \noindent
\newline
\clearpage
\begin{figure}[htp]
	\centering
\scalebox{0.60}{\includegraphics{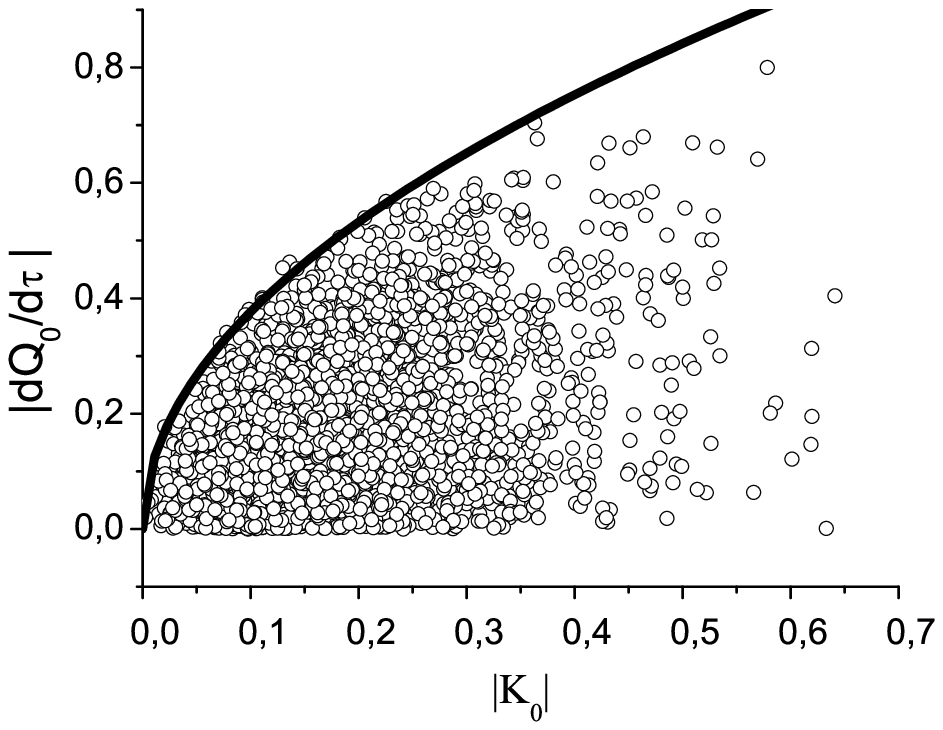}}
\scalebox{0.60}{\includegraphics{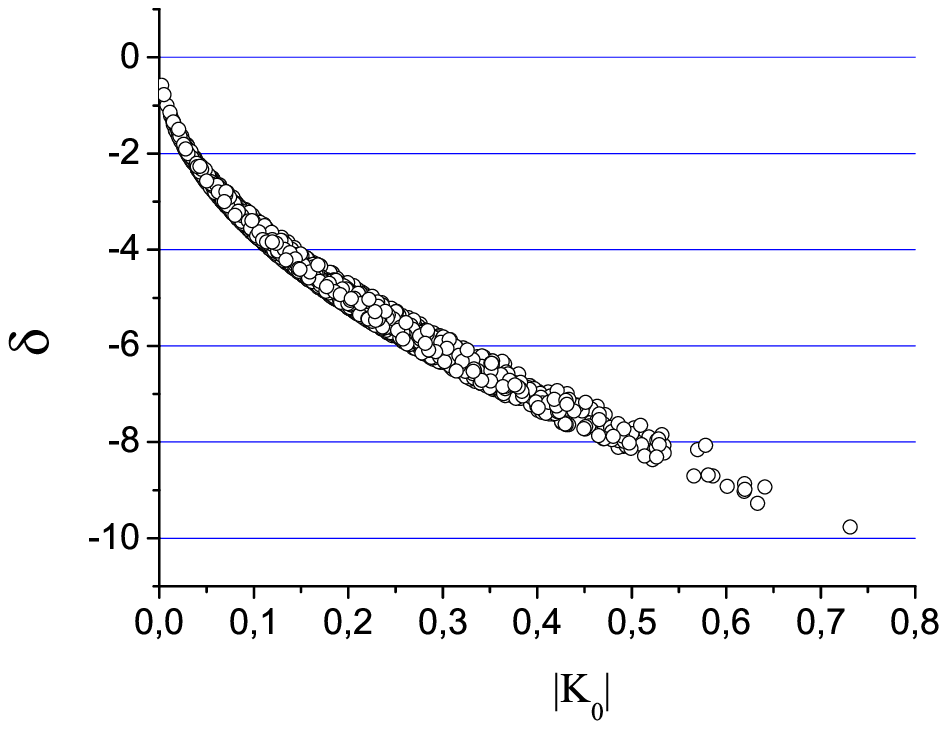}}
	\caption{Model B: 2 qubits. Numerical test of (26). Left : distribution of 3000 random points. Right: dependence of the saturation index on curvature  }
	\label{fig:model Bn2}
\end{figure}
\noindent
\newline
\begin{figure}[htp]
	\centering
\scalebox{0.60}{\includegraphics{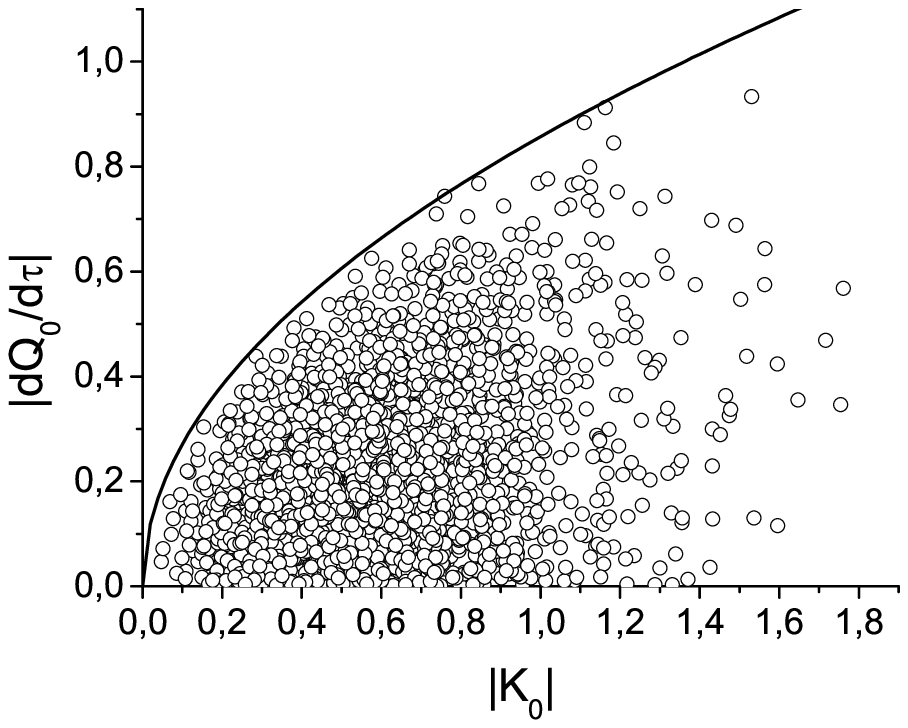}}
\scalebox{0.60}{\includegraphics{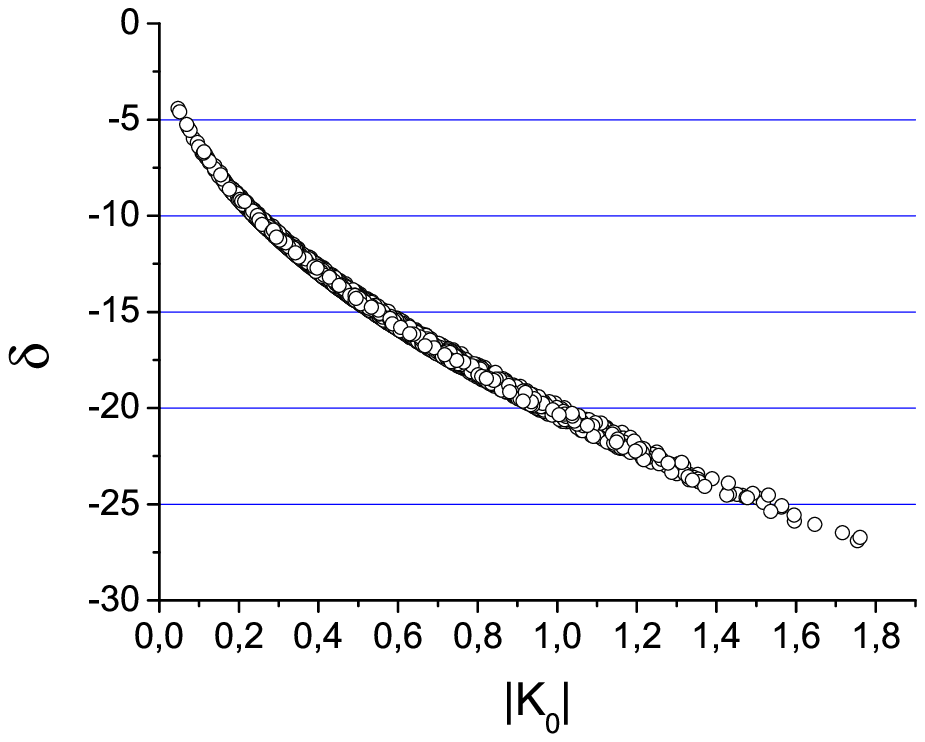}}
	\caption{Model B: 3 qubits. Numerical test of (26). Left : distribution of 3000 random points. Right: dependence of the saturation index on curvature  }
	\label{fig:model Bn3}
\end{figure}
  
 \noindent
\newline
\vspace{1 cm}\\
\begin{figure}[htp]
	\centering
\scalebox{0.60}{\includegraphics{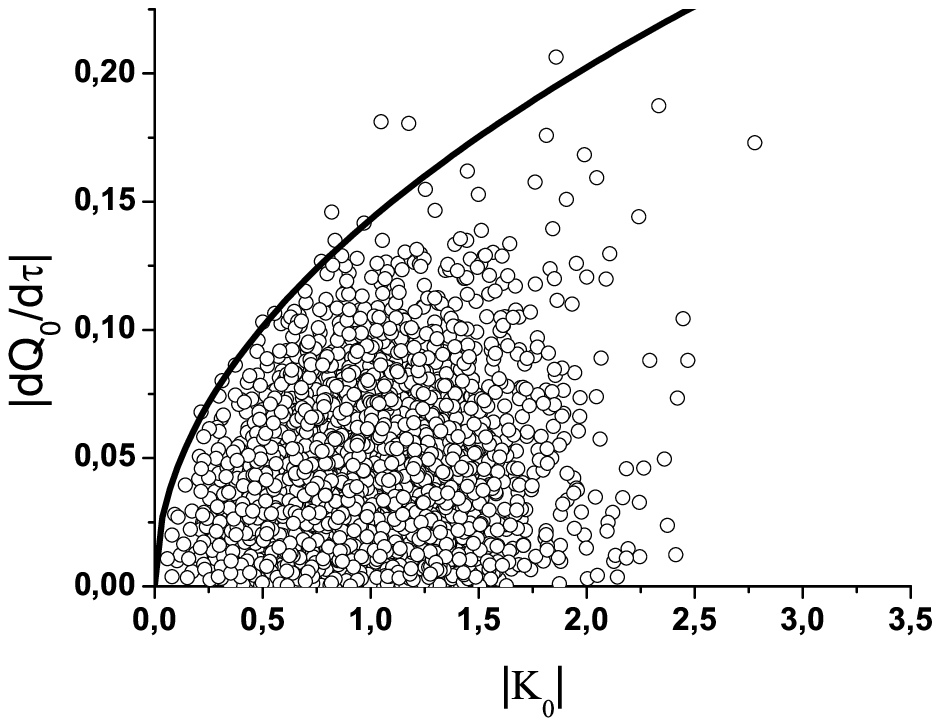}}
\scalebox{0.60}{\includegraphics{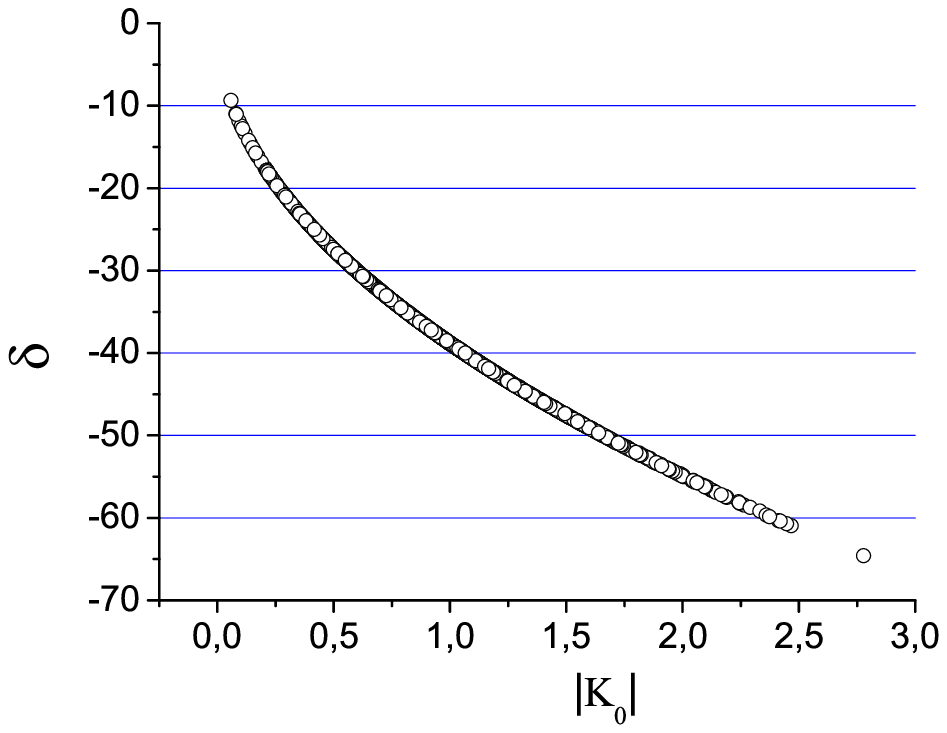}}
	\caption{Model B: 6 qubits. Numerical test of (26). Left : distribution of 3000 random points. Right: dependence of the saturation index on curvature  }
	\label{fig:model Bn6}
\end{figure}
 \noindent
\newline
\vspace{1 cm}\\
\begin{figure}[htp]
	\centering
\scalebox{0.60}{\includegraphics{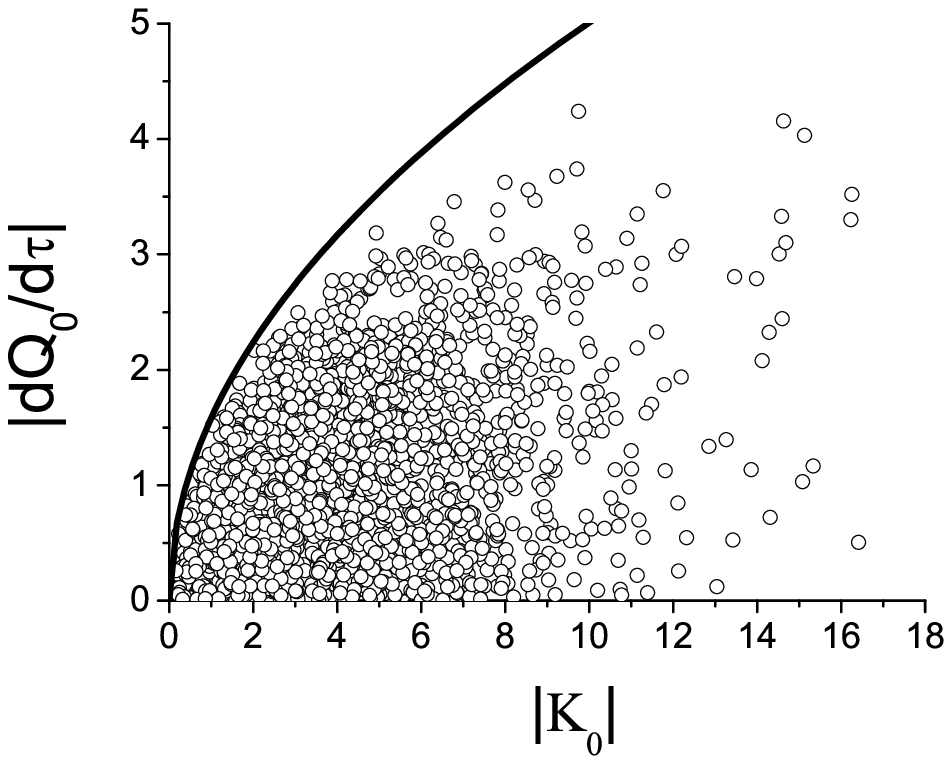}}
\scalebox{0.60}{\includegraphics{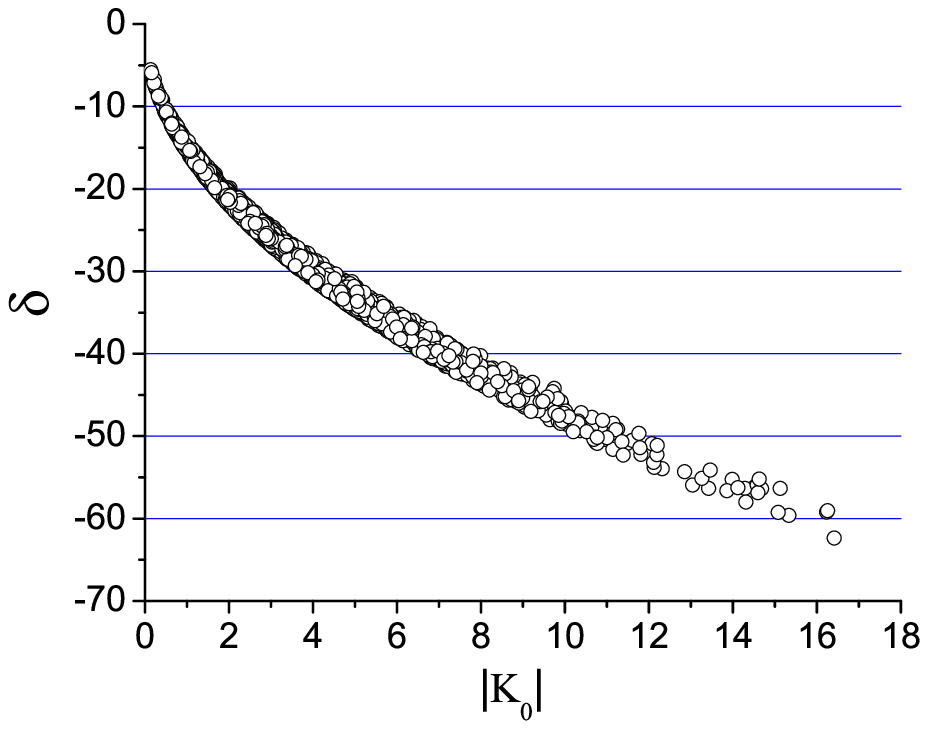}}
	\caption{Model C: 2 qubits. Numerical test of (26). Left : distribution of 3000 random points. Right: dependence of the saturation index on curvature. GUE perturbation  }
	\label{fig:model Cgue}
\end{figure}
\noindent
\newline
\vspace{1 cm}\\
\begin{figure}[htp]
	\centering
\scalebox{0.60}{\includegraphics{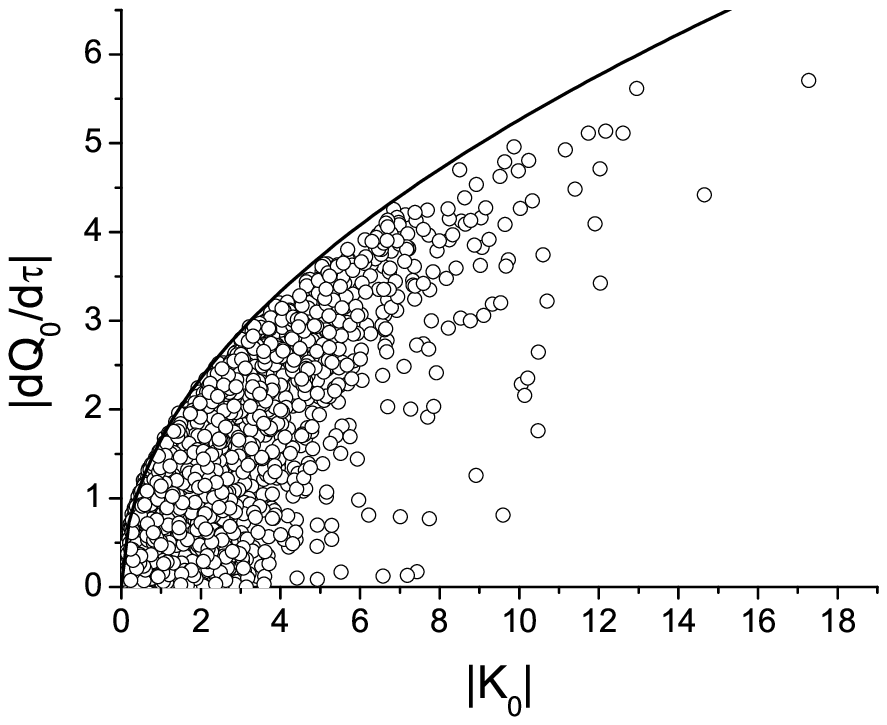}}
\scalebox{0.60}{\includegraphics{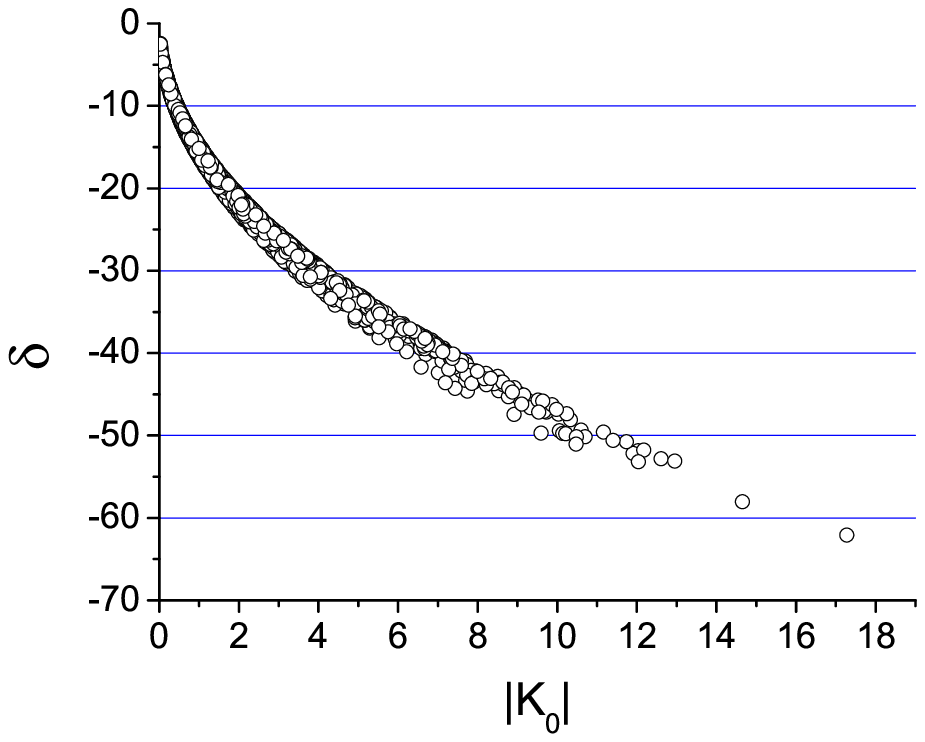}}
	\caption{Model C: 2 qubits. Numerical test of (26). Left : distribution of 3000 random points. Right: dependence of the saturation index on curvature. GOE perturbation.  }
	\label{fig:model Cgoe}
\end{figure}
 \noindent
\newline
\vspace{1 cm}\\
\indent
These results seem to confirm that the inequality is satisfied strictly.  The plots of $|\frac{\partial Q^{0}}{\partial\tau}|$ vs $|K_{0}|$ for all cases make it obvious that the inequality is certainly not saturated, but the plots of $\delta $ (the saturation index) vs $|K_{0}|$ gives us an indication how this lack of saturation depends on dynamics. It is interesting that the higher the value of $|K_{0}|$ the stricter the bounding of the entanglement change is. Of course for any specific model, both factors of the r.h.s of the inequality , namely b and $\sqrt{|K_{0}|}$ must be taken into account, and for comparison of the models we must , somehow, normalize this dependence. We report elsewhere on this point. 
 \vspace{1 cm}\\
\indent
\,6.2\, Dependence of the upper bound on the degree of chaos.
\vspace{1 cm}\\
\indent
We used two models to investigate how the parameter b of the upper bound of the proved inequality depends on the dynamical properties of the system with bi-partite entaglement. These models have different physical interpretation, one being a general RMT picture of chaotic systems, and the other a simplified spin-chain with defects. As it turned out, for both systems, the upper bound parametr b strongly depends on the degree of chaos in the dynamics of the system. This degree is quantified by the widely used parameter $\gamma $. Certainly more measures should be used, but we feel that , as a first step, the present results are very characteristic. We present the models, some numerical details and the results with some comments.
\indent
\newline
\newline
\indent
Model D
\newline
\indent 
The Hamiltonian of the model is the weighted sum of two random matrices, one with Poisson level statistics and the other with GOE \cite{guhr}. 
\begin{equation}
H(\theta)\,=\,cos(\theta)H_{P}+sin(\theta)H_{W}
\end{equation}  
We have chosen the dimension of the matrices to be $2^{7}\times 2^{7}$ to correspond to a system of 7 qubits. 
For $\theta=0$ we have regular behavior and for $\theta=\pi /2$ we have chaos. Thus we may vary $\theta$ from 0 to $\pi /2$ and check through the level spacing distribution parameter $\gamma$ the variation from the regular to the chaotic regimes. For each value of $\theta$ we used a 100 different random realizations of the Hamiltonian and calculated the averages $<b>$ and $<\gamma>$. We took a special care of the data to deal with the outliers, applying a small variation of the methods in \cite{santos}. To compute $\gamma$ we had to fit the data appropriately. We have chosen to do that using the Weibull distribution. 
\begin{equation}
y_{a,c}(x)\,=\,acx^{c-1}exp(-ax^{c})\,\,\,,\,\,\,x\in [0,+\infty]
\end{equation} 
The Weibull distribution contains as a special case the Brody distribution, which is usually applied to interpolate between the Poisson and the Wigner distributions. We use this model to see the dependence of the bound b on the degree of chaos. The model does not offer us the posibility to calculate the mean bi-partite entanglement, but it gives us the general behavior of b w.r.t. the underlying dynamics. In Figures 7, 8 and 9 we plot respectively the dependence of $\gamma$ and d on $\theta$, and of b on $\gamma$. The first gives the picture of the change of the behavior of the system from regularity, high $\gamma$ , to chaoticity , small $\gamma$. The other plots give the dependence of the bound on degree of regularity or chaoticity of the system. It is obvious that b takes its smallest values in the region of regularity. This means that the inequality is bounded more strictly, that is, that the entanglement change is more restricted when the system is more regular. 
\begin{figure}[htp]
	\centering
\scalebox{0.60}{\includegraphics{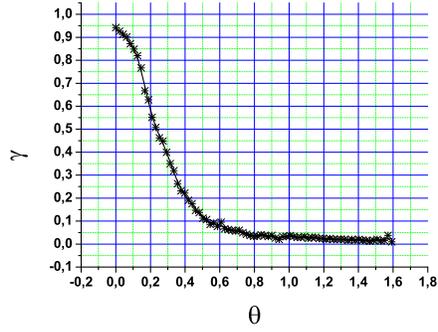}}
	\caption{Model D: 7 qubits. Dependence of the level statistics parameter $\gamma$ on the dynamical parameter $\theta$. The diagram shows the variation from regularity to chaos as a function of the critical parameter. }
	\label{fig:model D_1}
\end{figure}
\noindent
\newline
\vspace{1 cm}\\
\indent
\begin{figure}[htp]
	\centering
\scalebox{0.60}{\includegraphics{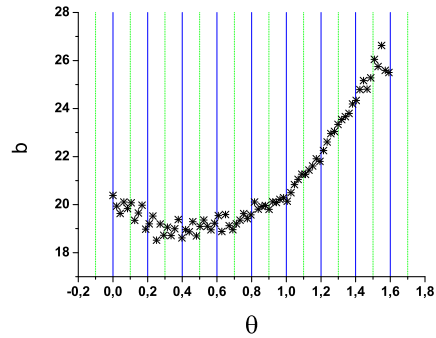}}
	\caption{Model D: 7 qubits. Dependence of the bound parameter b on the dynamical parameter $\theta$. The diagram shows the variation of the bound of the inequality  as a function of the critical parameter. }
	\label{fig:model D_2}
\end{figure}
\newline
\begin{figure}[htp]
	\centering
\scalebox{0.60}{\includegraphics{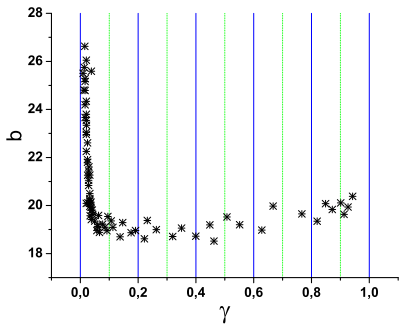}}
	\caption{Model D: 7 qubits. Dependence of the bound parameter b on  the level statistics parameter $\gamma$ . The diagram shows the variation of the bound of the inequality as a function of the measure of chaos. }
	\label{fig:model D_3}
\end{figure}
\indent
Model E
\newline
\indent
We use this model to test the dependence on the dynamics of both the bound parameter b and the mean bi-partite entanglement. This gives us the picture of how strict the inequality becomes and in the same time how much the entanglement is influenced by the underlying dynamics of the system. This model is well studied in the literature,  \cite{santos}. It describes a one-dimensional spin chain with nearest neighboor interaction with strength J. These qubits are coupled to an external homogeneous magnetic field. The system is perturbed by a distribution of random defects in the form of random interaction energies $h_{j}$ with the external field, with normal distribution
\begin{equation}
<h_{j}>\,=\,0\,\,\,,\,\,\,<h_{i}h_{j}>\,=\,d^{2}\delta_{ij}
\end{equation}   
\begin{equation}
H\,=\,\sum_{j=1}^{N}(h+h_{j})\sigma_{zj}\,+\,\frac{J}{4}\sum_{j=1}^{N-1}\vec{\sigma_{i}}\cdot\vec{\sigma_{j}}
\end{equation} 
It is known \cite{santos}, that the system is chaotic for the the values of d in the interval $0.1\leq d \leq0.5$ for J=1. We have analyzed this system for the interval  $0\leq d \leq 2.5$ for a chain of 9 qbits. For each value of d we used 100 random matrices to compute the averages $<Q>$, $<b>$ and $<\gamma>$, again using the Weibull distribution to fit level distribution density. The results are presented in the Figures 10,11 and 12. In the two parts of Figure 10 we plot $\gamma$ and Q vs the critical parameter d. In Figure 11 we plot b vs d and $\gamma$, and in Figure 12 we plot Q vs $\gamma$ and b. Figure 10 exhibits how the dynamical parameter d influences both the appearance of chaotic behavior and the establishing of entanglement. Figure 11 shows the dependence of the bound parameter b on the critical dynamical parameter and on the measure of chaos in the system, and Figure 12 gives the dependence of mean bi-partite entanglement on the measure of chaos and on the bound of the rate of change. From Figure 10 we see that at about the value of d=0.25, $\gamma$ is minimal and Q is maximal. This means that at this value we are closest to the chaotic behavior while in the mean time the mean bi-partite entanglement is maximal, something that conforms to similar results in the literature. From Figure 11 it follows that at this value of d the bound parameter b is maximal, something which means, that while the entanglement becomes maximal, the inequality bound parameter becomes less strict. Or, in other words, that the entanglement is more fragile. Finally Figure 12 reveals the direct relation between entanglement and the bound parameter b. It shows that more entanglement is associated to less strict bound , or to a more fragile regime. Of course, in order to draw conlusive results, we need  to use bigger matrices and certainly more realistic systems. But , even at this stage, our results point to the usefullness of our inequality for giving a general picture of the situations where we have the interplay of classical chaotic dynamics and the quantum regime of entanglement.     
\begin{figure}[htp]
	\centering
\scalebox{0.60}{\includegraphics{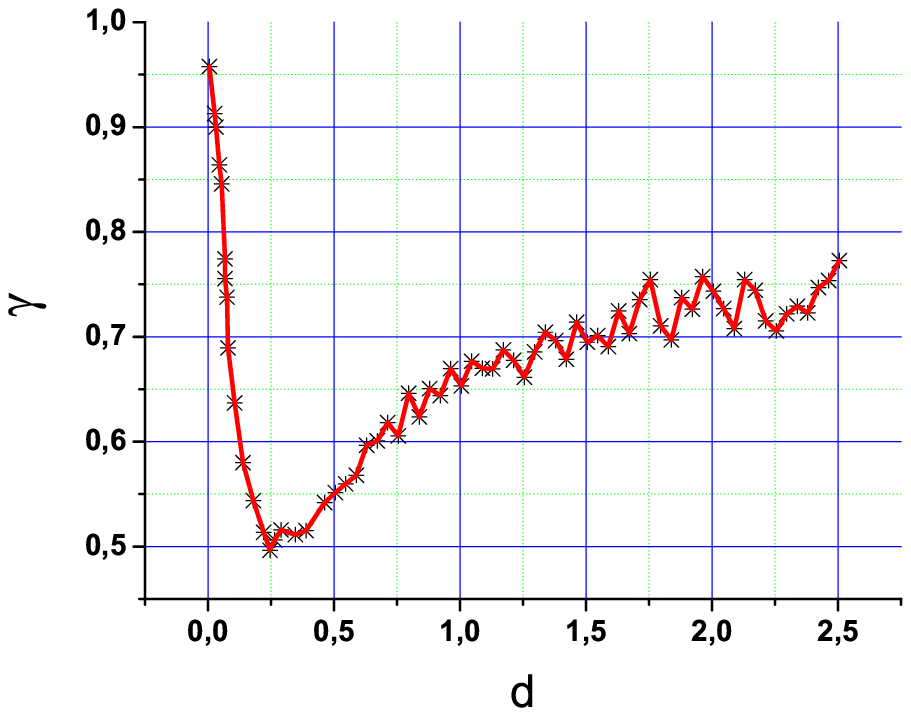}}
\scalebox{0.60}{\includegraphics{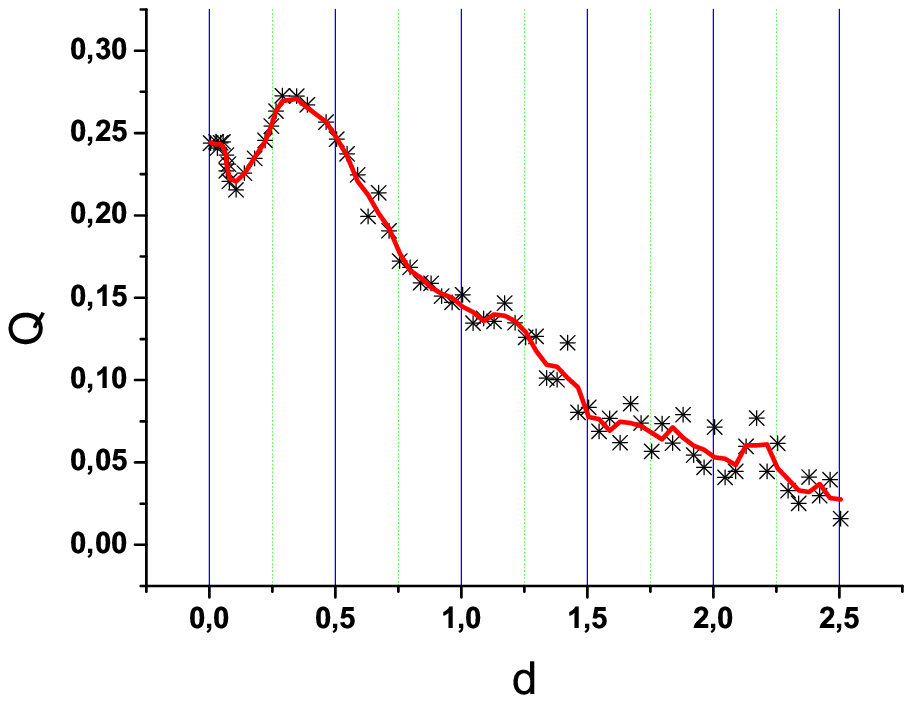}}
	\caption{Model E: 9 qubits. Dependence of the level statistics parameter and mean bi-partite entanglement on the dynamical parameter d  }
	\label{fig:model E_1ab}
\end{figure}
 \noindent
\newline
\begin{figure}[htp]
	\centering
\scalebox{0.60}{\includegraphics{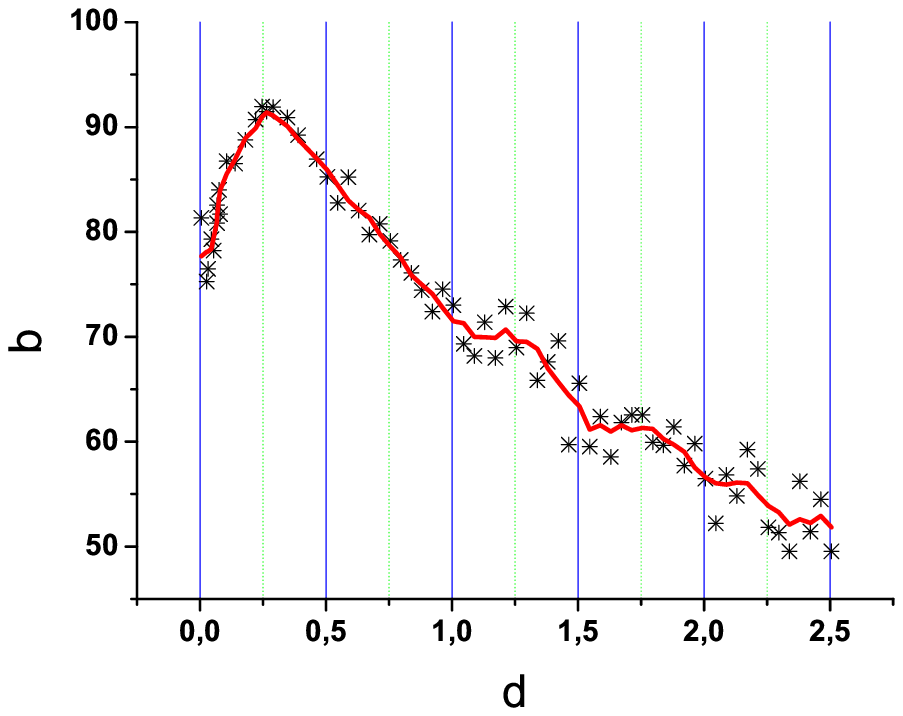}}
\scalebox{0.60}{\includegraphics{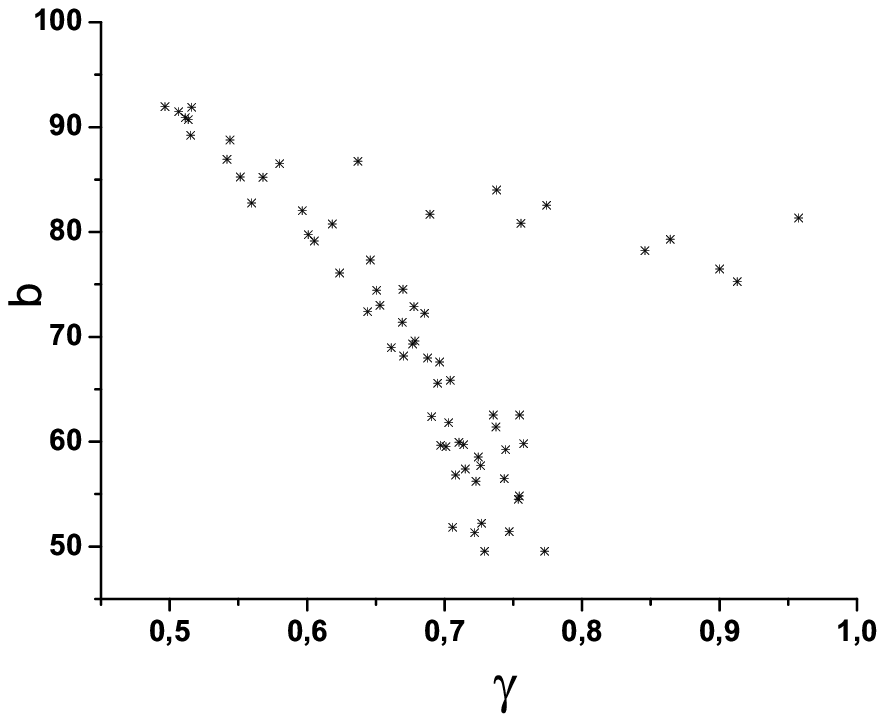}}
	\caption{Model E: 9 qubits. Dependence of the bound parameter b on the dynamical parameter d and on the measure of chaos }
	\label{fig:model E_2ab}
\end{figure}
 \noindent
\newline
\vspace{1 cm}\\
\begin{figure}[htp]
	\centering
\scalebox{0.60}{\includegraphics{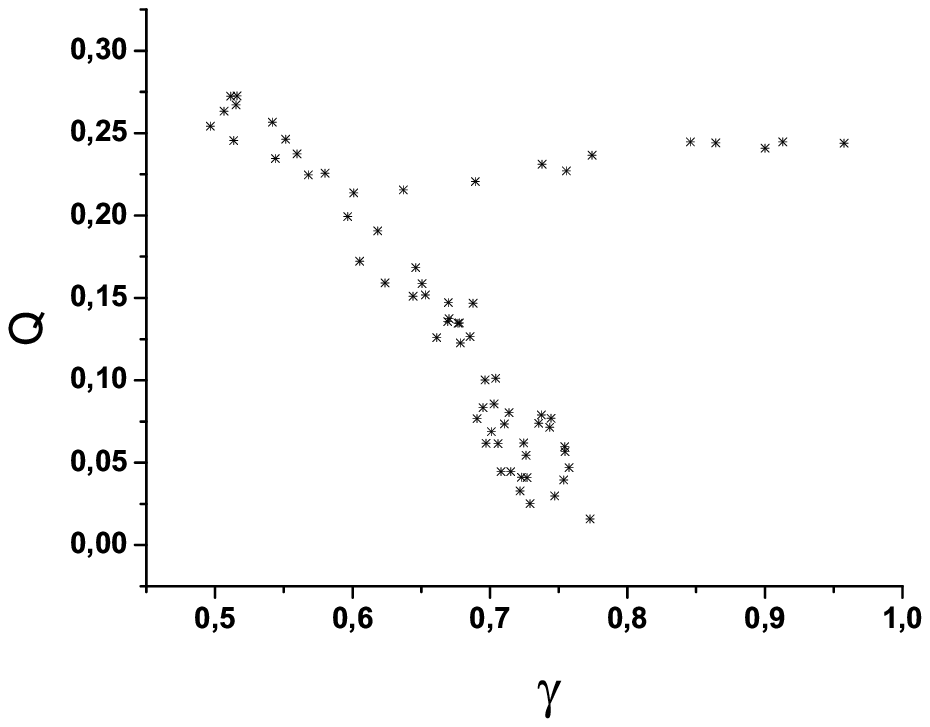}}
\scalebox{0.60}{\includegraphics{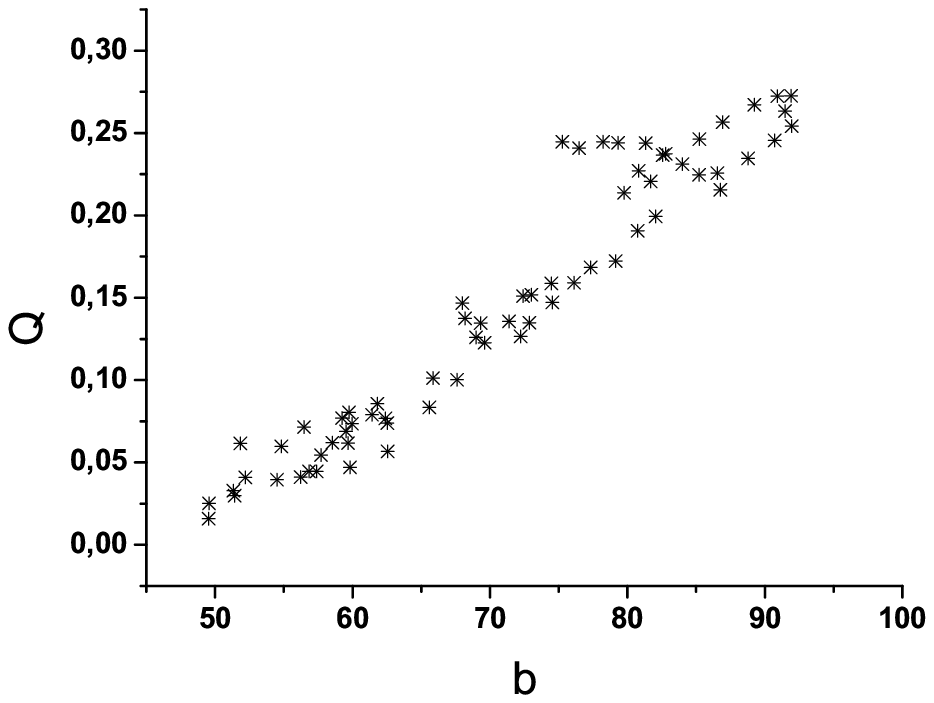}}
	\caption{Model E: 9 qubits. Dependence of the mean bi-partite entanglement on the measure of chaos and on the bound parameter b }
	\label{fig:model E_3ab}
\end{figure}
 \noindent
\newline
\indent
\newline
\indent
\,7.\, DISCUSSION.
\vspace{1 cm}\\
\indent 
 We have established a level-curvature dependent bound for the change of mean bi-partite entanglement under small changes of perturbation of the system. Extensive numerical tests were performed indicating no violation of this bound for a selected class of models. This bound depends on two factors, the parameter b and the curvature of the levels. Both factors reflect general properties of the undelying dynamics and combine to bound the rate of change of the mean bi-partite entanglement under small variations of a critical dynamical parameter. Using five  models we made extensive numerical investigation to check various aspects of the proposed inequality. First we tested the inequality with three different models and we found no violation.  The plots of $|\frac{\partial Q^{0}}{\partial\tau}|$ versus $|K_{0}|$ indicate that though the inequality is satisfied, it is far from being tight. To test the dependence of this lack of tightness , we plotted a saturation index vs  $|K_{0}|$. From these plots it is evident that the saturation is better with increasing mean curvature. The second type of tests concern the dependence of the bound parameter b on the dynamical property of regularity or chaos of the system. For two different models we first checked the dependence of the level statistic parameter $\gamma$ on the critical parameter of the models. We use $\gamma$ as a measure of the distance of the model from being integrable or chaotic. Then we plotted the parameter b of the bound vs both $\gamma$ and the critical dynamical parameter. It is evident from these plots that there is a strong dependence of the bound on the degree of chaos. Since b is computed for the unperturbed system, the size of the bound reflects somehow the fragility of entanglement under small perturbation of the values of the dynamical parameters. Finally the plots of the dependence of entanglement vs the measure of chaos and the bound parameter, give a first picture of how the proposed inequality could be used to approach the fragility of entanglement. The models we analyzed are general enough, and to our opinion, cover the properties of many specific models used in the literature. Our results, we believe, would be very useful for the study of the influence of inaccuracies of perturbation parameters or the influence of stochastic environments. Furthermore, using the curvature, one could design optimal strategies for entanglement and protection from decoherence. We currently investigate the use of the bound for models of concrete realization of Q.I.D.'s.

\end{document}